\documentclass[twocolumn,aps,prl,superscriptaddress]{revtex4-2}
\usepackage{color}
\usepackage{mathrsfs}
\usepackage{bm}
\usepackage{amsmath}
\usepackage{amssymb}
\usepackage{graphicx}
\usepackage{hyperref}
\usepackage{color}
\usepackage{siunitx}

\makeatletter
\hypersetup{hypertex=true, colorlinks=true, linkcolor=blue, anchorcolor=blue,citecolor=blue}

\usepackage{dcolumn}
\usepackage{bm}
\usepackage{mathrsfs}
\usepackage{amsfonts}
\usepackage{xcolor}
\usepackage{epstopdf}

\begin{document}
\title{Quantum Geometric Effects on the Higgs Mode in Flat-band Superconductors}
\author{Yuhang Xiao}
\affiliation{Anhui Province Key Laboratory of Low-Energy Quantum Materials and
Devices, High Magnetic Field Laboratory, HFIPS, Chinese Academy
of Sciences, Hefei, Anhui 230031, China}
\affiliation{Science Island Branch of Graduate School, University of Science and
	Technology of China, Hefei, Anhui 230026, China}
\author{Ning Hao}
\email{haon@hmfl.ac.cn}
\affiliation{Anhui Province Key Laboratory of Low-Energy Quantum Materials and
Devices, High Magnetic Field Laboratory, HFIPS, Chinese Academy
of Sciences, Hefei, Anhui 230031, China}

\begin{abstract}
Flat-band systems are of great interest due to their strong electron correlations and unique band geometry. Recent studies have linked the properties of Cooper pairs in flat-band superconductors to the quantum metric. Unlike prior studies primarily focused on quasiparticle fluctuations, in this work,  we investigate quantum geometric effects on a collective mode of the order parameter---the Higgs mode. We derive the quantum goemetric Higgs-mode correlation length and investigate whether the nonlinear electromagnetic response of the Higgs mode persists in flat band. It turns out that the quantum metric will replaces energy derivatives, playing a crucial role in third-harmonic generation (THG). We further numerically calculated the correlation length and THG in the Lieb lattice. In contrast to traditional single-band results, Higgs mode fluctuations contribute almost entirely to THG, with the quasiparticle contribution being negligible. This finding provides valuable guidance for detecting Higgs modes in flat-band systems through optical methods and reveals the profound influence of quantum geometry in flat-band systems.
\end{abstract}
\maketitle

\textit{Introduction}.--- Flat-band materials have garnered significant attention from both experimentalists and theorists due to their tendency to exhibit strong electron correlation effects and nontrivial band geometry\cite{kopnin2011high,uchoa2013superconducting,gulacsi2010route,shaginyan2010scaling,TBG1,TBG2,TBG3,TBG4,TBG5,TBG6,TBG7}. Studies on flat-band superconductivity have revealed that the properties of Cooper pairs are closely related to the geometric characteristics of the normal-state electronic bands. Specifically, in flat-band systems, the effective mass and stiffness of Cooper pairs are defined by the quantum metric\cite{peotta2015superfluidity,torma2022superconductivity,liang2017band,CooperPairEffectiveMass,FlatBandExp1}. Over the past few decades, research on band geometry has primarily focused on global topological properties to classify topological materials\cite{ryu2010topological,zhang2001four}. The key concepts are the Berry curvature and the Chern number, obtained by integrating the curvature over a closed manifold. Unlike the Chern number, the quantum metric is a local geometric quantity that allows us to define a notion of distance on the manifold in which the quantum states reside\cite{QGT1,QGT2,QGT3,QMinChiralLattice}.

In the superconducting condensate, there are essentially three types of excitation modes: quasiparticle excitations, phase (Goldstone) and amplitude(Higgs) modes of the order parameter. The phase mode arises from the spontaneous breaking of the global U(1) symmetry in superconductors. As a Goldstone mode, it is gapless in neutral systems. But in charged systems, it couples to the gauge field, acquiring a gap corresponding to the plasma frequency\cite{higgs1964broken1,higgs1964broken2,guralnik1964global}. In contrast, the Higgs mode is a gapped excitation, with its gap comparable to the quasiparticle gap\cite{tsuji2015theory,littlewood1982amplitude}. Its spectrum extends into the quasiparticle continuum, which makes it challenging to distinguish the Higgs mode from the quasiparticle background. Fortunately, the nonlinear coupling between the Higgs mode and the electromagnetic field, especially the third-harmonic generation (THG)\cite{matsunaga2014light,tsuji2016nonlinear,schwarz2020theory,murotani2017theory,HiggsExp1,HiggsExp2,HiggsExp3,HiggsExp4,HiggsExp5}, has enabled experimentalists to probe its effects using THz optical techniques.

Previous studies on quantum geometric effects in flat-band superconductors mainly concentrated on the contribution of quasiparticle fluctuations\cite{QGPairPotential,GeoStiffnessofCompositeBands}. By deriving the Ginzburg-Landau effective action from microscopic theory and performing a gradient expansion, it becomes evident that the quantum metric appears in the velocity terms of the superconducting order parameter\cite{GeoStiffnessGLTheory}. In our work, we follow this framework to calculate the Higgs-mode correlation length. Furthermore, we answer an important question: does the nonlinear electromagnetic response of the Higgs mode persist in flat-band systems? In single-band systems, if we take the flat-band limit, THG will vanish entirely because all effects involve energy derivatives. However, we find that when projecting a multiband system onto a flat band, the quantum metric replaces the energy derivatives, thereby contributing fully to the electromagnetic response. Finally, we calculate the correlation length and the third-harmonic current using a specific lattice model, the Lieb lattice. The results indicate that the contribution from the Higgs mode is nearly $10^5$ times greater than that of quasiparticle fluctuations. Therefore, any THG observed in flat-band systems can be attributed almost entirely to interactions mediated by the Higgs mode, with the quantum metric playing a crucial role at the interaction vertex.

\textit{Effective theory}.--- To illustrate the quantum geometric contributions to the Higgs-mode correlation length, we begin with a microscopic model $H=H_{0}+H_{int}$ where $H_{0}$ is a generic non-interacting Hamiltonian with time-reversal symmetry and $H_{int}$ is the nearest-neighbor antiferromagnetic interaction (exchange coupling strength $J>0$) defined on the square lattice.
\begin{eqnarray}
    &&H_{0}=\sum_{\bm{k} \sigma} \sum_{\alpha \beta} c_{\alpha \sigma}^{\dagger}(\bm{k}) \mathcal{H}_{\alpha \beta}^{\sigma}\left(\bm{k}\right) c_{\beta \sigma}(\bm{k}),\label{H_0}\\
    &&H_{int}=J \sum_{\langle i j\rangle} \bm{S}_{i} \cdot \bm{S}_{j}.\label{H_int}
\end{eqnarray}
Here, $\sigma\in\{\uparrow,\downarrow\}$, $\alpha(\beta)=1,...,N_{b}$ and $i(j)$ lable the spin, orbital(or sublattice) and site degrees of freedom of electrons, respectively. Among the energy bands that the non-interaction Hamiltonian produces, there is a non-degenerate isolated flat one with a gap ($E_g$) separating it from the other dispersive bands. In the interaction, $\bm{S}_{i}=\frac{1}{2}\textstyle\sum_{\sigma \delta} \sum_{\alpha} c_{i \alpha \sigma}^{\dagger}\bm{\sigma}_{\sigma \delta} c_{i \alpha \delta}$ is the spin operator of electrons. We choose this interaction to incorporate both s-wave and d-wave pairing, and we would like to emphasize that our results are not heavily dependent on the specific microscopic model. 

The partition function of this model can be written by the path-integral formalism $\mathbin{Z}=\int\mathcal{D}( c^{\dagger},c)exp(-S) $ wherein the Euclidean action reads $S=\int_{0}^{\beta}\mathrm{d}\tau(c^{\dagger}\partial_{\tau}c+H(\tau))$ with $\beta=1/T$. After carrying out the standard Hubbard-Stratonovich transformation\cite{HStrans,coleman} to introduce the bosonic order-parameter field $\Delta_{\alpha\beta}(q)$, the interaction term in the action yields $ S_{int}=-\beta \sum_{k q} \Delta_{\alpha \beta}(q) \zeta_{\bm{k}, \bm{k}-\bm{q}} c_{\alpha \uparrow}^{\dagger}(k) c_{\beta \downarrow}^{\dagger}(-k+q)+\text{H.c.}$, where the unbolded momentum indeices stand for the four-momentum consisting space and frequency components, and the sum over the orbital indices is implicit. The interaction given by Eq.(\ref{H_int}) gives rise to both s-wave and d-wave pairing; however, in practice, only one pairing state is typically the true ground state. Thus we do not consider the coexistence of both but instead a priori assume the existence of one possible state. The form factors $\zeta^{s/d}_{\bm{k},\bm{k}^{\prime}}$ for extended s-wave and d-wave pairing are $\frac{1}{2} \left ( \cos k_{x}\pm\cos k_{y}\right ) +(\bm{k}\to\bm{k}^{\prime} ) $, respectively. We also define $\zeta^{s/d}_{\bm{k}}\equiv\zeta^{s/d}_{\bm{k},\bm{k}}$ for convenience in the following discussion. We will omit the 
$s/d$ superscripts for now and specify them when needed later.

By defining the projected flat-band operator $c_{\sigma}^{\dagger}(\bm{k}) \equiv \sum_{\alpha} u_{\alpha}^{\sigma}(\bm{k}) c_{\alpha \sigma}^{\dagger}(\bm{k})$ with $u_{\alpha}^{\sigma}(\bm{k})$ representing the flat-band Bloch wave function, we construct an effective theory for the superconductivity in the flat band. This theory is only applicable when the band gap $E_g$ greatly exceeds the saddle-point value of $\Delta(q)$ (i.e. the gap of superconductors), which is known as the isolated-flat-band condition. The action can be written in the Bogoliubov-de-Gennes (BdG) form and the inverse of Nambu propagater $\mathcal{G}(k,k^{\prime})$ is given by
\begin{eqnarray}
    \mathcal{G}^{-1}=&&i\omega_{n}\tau_{0}-\xi_{\bm{k}}\tau_{3}\nonumber\\
    &&+\operatorname{Re}\left[\Delta\left(k-k^{\prime}\right) f(\bm{k},\bm{k}-\bm{k}^{\prime})\right] \zeta_{\bm{k},\bm{k}^{\prime}}\tau_{1}\nonumber\\
    &&-\operatorname{Im}\left[\Delta\left(k-k^{\prime}\right) f(\bm{k},\bm{k}-\bm{k}^{\prime})\right] \zeta_{\bm{k},\bm{k}^{\prime}}\tau_{2},
\label{NambuPropagator1,mt}
\end{eqnarray}
where $\xi_{\bm{k}}$ is the band dispersion relative to the chemical potential, and the factor $f(\bm{k},\bm{q})\equiv\sum_{\alpha} u_{\alpha}^{*}(\bm{k}) u_{\alpha}(\bm{k}-\bm{q})=\langle m, \bm{k} \mid m,\bm{k}-\bm{q}\rangle$ containing the wave-function imformations of the flat band. We have introduced the notations $\xi (\bm{k}) \equiv \xi_{\uparrow}(\bm{k})=\xi_{\downarrow}(-\bm{k})$ and $u_{\alpha}(\bm{k}) \equiv u_{\alpha}^{\uparrow}(\bm{k})=u_{\alpha}^{\downarrow \ast}(-\bm{k})$ to account for time-reversal symmetry. Additionally, the uniform-pairing condition $\Delta_{\alpha\beta}(q)=\Delta(q)\delta_{\alpha\beta}$ is adopted, which implying that there is no inter-orbital pairing and that the pairing strength remains the same for all orbitals. This is a crucial condition to give the quantum geometric effects in the context of superconductivity\cite{CooperPairEffectiveMass,GeoStiffness}. If it is not met, a pair-density-wave instabililty may occur \cite{PDWfromQG}. Detailed derivation for the effective action can be found in the Supplemental Material\footnote{Supplementary materials}, and similar formalism can also be found in Refs. \cite{GeoStiffnessGLTheory,GeoEffectGLTheory}. 

All kinds of Fluctuation beyond the mean-field description are encoded in the inverse propagator $\mathcal{G}^{-1}$. They can be derived by expanding the order-parameter field about its saddle point. We take a uniform saddle point $\Delta_0$ and write the fluctuations of the amplitude(Higgs) and phase(Goldstone) mode as $\Delta(x)=\Delta_{0}(1+\rho(x)) e^{i \theta(x)} \simeq \Delta_{0}+\Delta_{0} \rho(x)+i \Delta_{0} \theta(x)$, where $\Delta_{0}$, $ \rho(x)$ and $\theta(x) \in \mathbb{R}$. And correspondingly $\Delta(k)=\Delta_{0} \delta_{k=0}+\Delta_{0} \rho(k)+i \Delta_{0} \theta(k)$. Inserting it into the Nambu propagator Eq.(\ref{NambuPropagator1,mt}), we obtain the second-order effective action 
\begin{equation}
    S_{2}=\beta \sum_{q}\left(\theta^{*}(q)\ \rho^{*}(q)\right)\left(\begin{array}{ll}
    \chi_{\theta \theta}(q) & \chi_{\theta \rho}(q) \\
    \chi_{\rho \theta}(q) & \chi_{\rho \rho}(q)
    \end{array}\right)\binom{\theta(q)}{\rho(q)}
\label{action2,mt}
\end{equation}
where the exact form of the susceptibility kernel $\hat{\chi}(q)$ can be found in the Supplemental Material. Taking the static (i.e. $iq_{n}=0$) value and expanding the susceptibilities to the second-order of momentum $\bm{q}$, we observe the emergence of the quantum geometric effect. To see it clearly, we extract the quantum-metric-related contributions within some miscellaneous terms. At zero temperature, they are given by
\begin{widetext}
\begin{align}
    \chi_{\theta \theta / \rho \rho}(\bm{q})&\simeq \frac{N_{b} \Delta_{0}^{2}}{J} \mp \frac{\Delta_{0}^{4}}{8} \sum_{\bm{k}} \frac{\zeta_{\bm{k}}^{4}}{E_{\bm{k}}^{3}}\left(1-q_{i} q_{j} g_{i j}(\bm{k})\right)
    -\frac{\Delta_{0}^{2}}{4} \sum_{\bm{k}} \frac{E_{\bm{k}}^{2}+\xi_{\bm{k}}^{2}}{E_{\bm{k}}^{3}} \zeta_{\bm{k}}^{2}\left(1-q_{i} q_{j} g_{i j}(\bm{k})\right),\label{chi_rhorho}\\
    \chi_{\theta \rho / \rho \theta}(\bm{q})&\simeq \frac{N_{b} \Delta_{0}^{2}}{J} \pm i \frac{\Delta_{0}^{2}}{4} \sum_{\bm{k}} \frac{E_{\bm{k}}^{2}+\xi_{\bm{k}}^{2}}{4 E_{\bm{k}}^{3}} \zeta_{\bm{k}}^{2}\left(1-q_{i} q_{j} g_{i j}(\bm{k})\right).\label{chi_thetarho}
\end{align}
\end{widetext}
Here $E_{\bm{k}}= (\xi_{\bm{k}}^{2}+\Delta_{\bm{k}}^{2})^{1/2}$ is the quasi-particle dispersion with $\Delta_{\bm{k}}=\Delta_{0} \zeta_{\bm{k}}$, and $g_{ij}(\bm{k})$ is the quantum metric. Generically speaking, $g_{ij}$ is a local geometric quantity in the Hilbert space which describes the difference between two adjacent normalized quantum states in the parameter manifold $\{\bm{\lambda}\}$. Its definition is given by $\left | \left \langle \bm{\lambda}  |\bm{\lambda}-\delta\bm{\lambda} \right \rangle  \right |^{2}\simeq 1-\sum_{i,j}\delta\lambda_{i}\delta\lambda_{j}g_{ij}(\bm{\lambda}) $ and can also be written specifically as $g_{i j}(\bm{\lambda})=\operatorname{Tr}\left[\partial_{\lambda_{i}} P(\bm{\lambda}) \partial_{\lambda_{j}} P(\bm{\lambda})\right]$\cite{SCinTMG} via the projection operator $P(\bm{\lambda})$. 

By setting $\xi_{\bm{k}}=0$, we apply the above expressions to the flat band.  The Higgs-mode susceptibility defined by $\chi_{H}=\chi_{\theta\theta}/\operatorname{det}\hat{\chi}=\chi_{\theta \theta}/(\chi_{\theta \theta} \chi_{\rho \rho}-\chi_{\theta \rho} \chi_{\rho \theta})$\cite{VisibilityofHiggsMode} takes the form of $\chi_{H}\propto 1+G_{i j} q_{i} q_{j}$. And the correlation length $\xi$ is given by $\xi^{2}=\sqrt{\operatorname{det} G_{i j}}$. The results for the cases of s-wave and d-wave pairing are as follows.:
\begin{equation}
    \xi_{s/d}^{2}=\sqrt{\frac{16 \pi^{6} x^{2}-22 \pi^{4} x+33 \pi^{2}}{8\left(\pi^{2} x-3\right)\left(16 \pi^{2} x-11\right)}} \bar{g}^{s/d}.
\label{xi^2}
\end{equation}
Here we defined a dimensionless factor $x\equiv N_{b}\Delta_{0}/J$ and average quantum metric in the Brillouin zone: \begin{equation}
 \bar{g}^{s/d}=\sqrt{\operatorname{det} \textstyle\sum_{\bm{k}} \left|\zeta^{s/d}_{\bm{k}}\right|g_{i j}(\bm{k})}.   
\end{equation}
We observe that the dependence between the correlation length and the interaction strength is manifested by the function of the dimensionless parameter $x$ as shown in Eq.(\ref{xi^2}) while the effects of quantum geometry appear in the weighted integral of the quantum metric. And the momentum structure of the weight function is determined by the pairing symmetry. With dynamic susceptibility ($iq_{n}\ne0$) taken into account, the quantum geometric correlation length will also impacts the quadratic term $\bm{q}^2/2m^*$ in the Higgs-mode spectrum. This effect will be clearly reflected in its contribution to the mass of the Higgs mode through $\xi^2\sim 1/m^*$.

\textit{Electromagnetic response}.--- The experimental detection of the Higgs mode in superconductors relies on optical methods. Thus, we now proceed to investigate the electromagnetic response of the Higgs mode and the role played by the quantum metric in this context. To grasp the main idea regarding the geometric effect, we make a slight adjustment to the microscopic model used above. Specifically, we now consider a model in which the center-of-mass momentum of Cooper pair is zero. The rationale behind this modification is that even without considering the momentum dependence of the order parameter, the presence of an electromagnetic field leads to the emergence of the quantum metric. We accomplish this by introducing the electromagnetic gauge field $\bm{A}$ through minimal coupling, i.e.$\mathcal{H}_{\alpha\beta}^{\sigma}(\bm{k})\to\mathcal{H}_{\alpha\beta}^{\sigma}(\bm{k}-\bm{A})$ (with $c=1$ and $e=1$). This transformation can also be realized in real space Hamiltonian through Peierls substitution. Similar to the case of non-zero center-of-mass pair momentum, the quantum metric plays a role by expanding the inner product $\left|\sum_{\alpha} u_{\alpha}^{*}(\bm{k}-\bm{A}) u_{\alpha}(\bm{k}+\bm{A})\right|=1-2\sum_{i, j} g_{i j}(\bm{k}) A_{i}(\tau) A_{j}(\tau)+o(A^{3})$. Furthermore, we expand $\xi_{\sigma}(\bm{k}-\bm{A})$ up to the second order of the guage field as $\xi_{\bm{k}-\bm{A}} =\xi_{\bm{k}}+\frac{1}{2} \partial^{2}_{ij} \xi_{\bm{k}} A_{i} A_{j}+o(A^{3})$, where the $A^{2}$ term corresponds to the diamagnetic current, and the first-order term which is the paramagnetic term vanishes since we also assumed the parity symmetry $\xi_{\bm{k}}=\xi_{-\bm{k}}$. The action for both the Higgs mode, electrons and guage field can be obtained (see  Supplemental Material), and the Nambu propagator now takes the following form:
\begin{eqnarray}
    \mathcal{G}^{-1}(\bm{k}, \tau)= && -\partial_{\tau} \tau_{0}-\xi_{\bm{k}} \tau_{3}+\Delta_{\bm{k}} \tau_{1}+\rho(\tau) \zeta_{\bm{k}} \tau_{1} \nonumber \\
   && -\frac{1}{2} \sum_{i j} \partial_{i j}^{2} \xi_{\bm{k}} A_{i}(\tau) A_{j}(\tau) \tau_{3}\nonumber \\
   && -2\Delta_{\bm{k}} \sum_{i j} g_{i j}(\bm{k}) A_{i}(\tau) A_{j}(\tau) \tau_{1}\nonumber  \\
   && -2\zeta_{\bm{k}} \sum_{i j} g_{i j}(\bm{k}) A_{i}(\tau) A_{j}(\tau) \rho(\tau) \tau_{1},
\label{NambuPropagator2,mt}
\end{eqnarray}
in which the first line is defined as $\mathcal{G}_{0}^{-1}(\bm{k},\tau)$ and the remaining terms are $-\Sigma(\bm{k},\tau)$, the self-energy. We do not consider the phase fluctuations (Goldstone mode) here, as they can be absorbed into the gauge field and pushed to the high-energy domain due to the well-known Anderson-Higgs mechanism\cite{HiggsMechanism1,HiggsMechanism2,HiggsMechanism3}. Comparing Eq.(\ref{NambuPropagator2,mt}) with the self-energy in the single band superconductors\cite{THG1,THG2}, there are two guage-field terms in $\tau_{1}$ channel induced by the quantum metric. Specially, the last term in Eq. (\ref{NambuPropagator2,mt}) indicates the coupling between the guage field and the Higgs mode, which is also absent in the single-band systems. We also find that the Higgs propagator $\mathcal{G}_{H}^{-1}\left(i \omega_{m}, i \omega_{n}\right)$ will take a rather complicated form where the guage field contributes via the effect of quantum geometry. The additional quantum-metric-related terms will significantly influence the dispersion of Higgs mode. But luckily, to derive the action for the guage field to the order of $\bm{A}^{4}$, we can neglect these terms leaving only the bare Higgs propagator, i.e.$\mathcal{G}_{H}\left(i \omega_{m}, i\omega_{n}\right)\simeq\mathcal{G}_{H}^{0}\left(i \omega_{m}\right)=\left(\frac{2}{J}+\sum_{\bm{k}} \zeta_{\bm{k}}^{2} \chi_{1 1}\left(\bm{k}, i \omega_{m}\right)\right)^{-1}$, 
in which $\chi_{11}\left(\bm{k}, i \omega_{m}\right)=\frac{1}{\beta} \sum_{i \omega_{n}} \operatorname{Tr}\left[\mathcal{G}_{0}\left(\bm{k}, i \omega_{n}\right) \tau_{1}\mathcal{G}_{0}\left(\bm{k}, i \omega_{n}+i \omega_{m}\right) \tau_{1}\right]$, corresponding to the fermion bubble. The full expression of the Higgs propagator can be found in the Supplemental Material. Integrating out the fermions and the Higgs mode, we obtain an effective electromagnetic action:
\begin{equation}
    S[A]= \int \mathrm{d} \omega  \Tilde{A}^{2}(-\omega) K(\omega) \Tilde{A}^{2}(\omega).
\label{action3,mt}
\end{equation}
Here we consider monochromatic polarized light, say, x-polarized light $\bm{A}(t)=A_{0}\hat{e}_{x}\cos (\Omega t)$, and define $A^{2}_{xx}(\omega)=\Tilde{A}^{2}(\omega)$. To avoid ambiguity, we clarify that the relation between $\Tilde{A}^{2}(\omega)$ and $A(\omega)$ is given by a convolution integral $\Tilde{A}^{2}(\omega)=\int\mathrm{d}\omega^{\prime}A(\omega^{\prime})A(\omega-\omega^{\prime})$. 
The quantum-geometric optical kernel in Eq.(\ref{action3,mt}) reads
\begin{eqnarray}
    K(\omega)= && \Delta_{0}^{2} \sum_{\bm{k}} g_{x x}^{2}(\bm{k}) \chi_{11}(\bm{k}, \omega) \nonumber\\
   &&-\Delta_{0}^{2} \mathcal{G}_{H}^{0}(\omega)\left[\sum_{\bm{k}} g_{x x}(\bm{k}) \chi_{11}(\bm{k}, \omega)\right]^{2}.
\label{OptKernel,mt}
\end{eqnarray}

\begin{figure}
\includegraphics[width=0.98\linewidth]{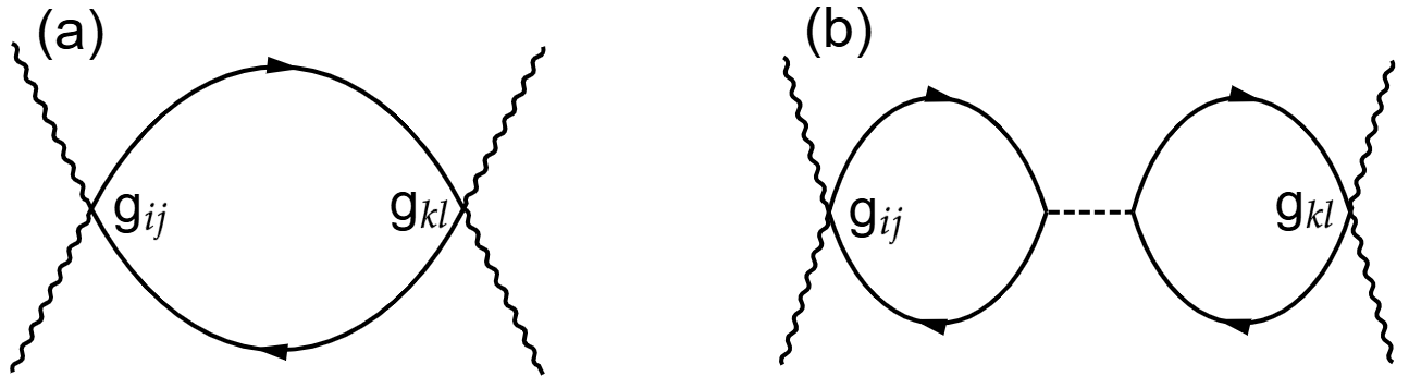}
\caption{\label{fig:FeyDiaforKernel} Feynman diagrams for the electromagnetic effective action. Wavy lines, solid lines and dashed lines stand for the guage field $A_{i}$, quasi-particles and Higgs mode, respectively. Two contributions in the quantum geometric optical kernel are diagrammatically shown in (a) $\chi_{11}$ term and (b) $\chi_{11}\mathcal{G}_{H}\chi_{11}$ term. The vertices in the diagrams are related to the quantum metric $g_{ij}$.}
\end{figure}
The first term in Eq. (\ref{OptKernel,mt}) gives the contribution of quasi-particle fluctuations and the second term gives the Higgs-mode mediated fluctuations. The quantum metric functions as the interaction vertices as shown in Fig. \ref{fig:FeyDiaforKernel}. The Higgs propagator and fermion bubble $\chi_{11}$ at zero temperature are calculated in the Supplemental Material. Via Eqs. (\ref{action3,mt}) and (\ref{OptKernel,mt}), the current can be evaluated by taking the functional derivative $j(t)=-\delta S/\delta A(t)$, and its third-harmonic component is given by the quantum geometric optical kernel as $j(3\Omega)\propto K(2\Omega)$.

\begin{figure}
\includegraphics[width=0.99\linewidth]{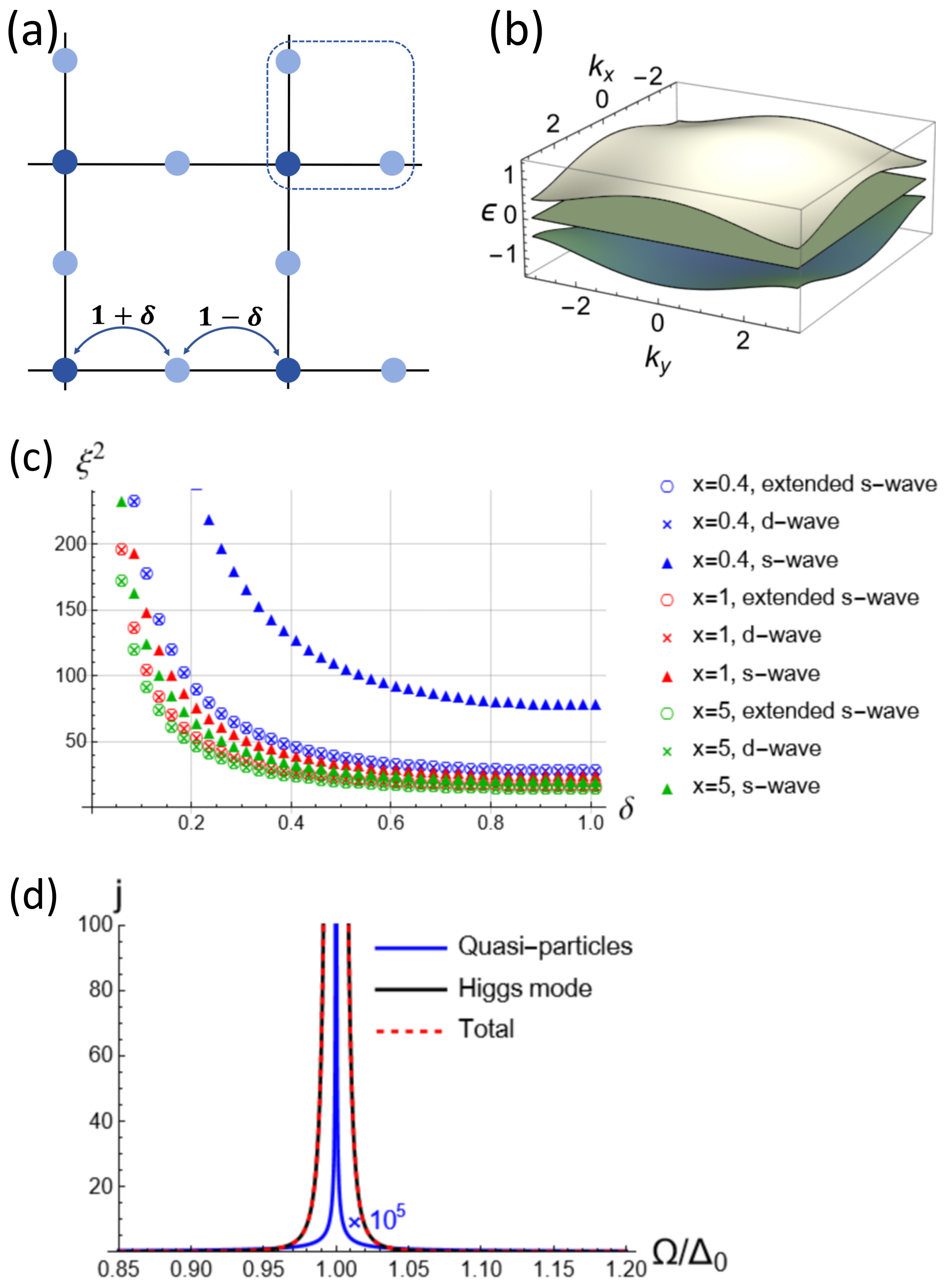}
\caption{\label{fig:Results}(a) Two dimensional Lieb lattice and its unit cell. The inter-unit-cell and intra-unit-cell hopping are $1-\delta$ and $1+\delta$, respectively. (b) The band dispersion for $\delta=1/3$. There is a robust flat band at zero energy. (c) The quantum geometric correlation length of Higgs mode for different pairing symmetry. (d) The third-harmonic current $j(3\Omega)$ as function of $\Omega$ ($\delta=1/3$ and $\Delta_{0}=1$) in the Lieb-lattice flat band.}
\end{figure}

\textit{Application to Lieb lattice}.--- We apply the above results to the two-dimensional Lieb lattice, which is a three-band system that features a robust flat band as shown in Fig. \ref{fig:Results}(b). In Fig. \ref{fig:Results}(c), the quantum geometric correlation length of this lattice model is calculated numerically using Eq. (\ref{xi^2}).  We observe that extended s-wave and d-wave pairing exhibit identical results. In the strongly correlated regime (i.e.$x$ is small), the differences between s-wave ($\zeta_{\bm{k}}=1$) and d-wave are pronounced, while in the weakly correlated regime, the difference gradually diminishes. We can also find that, as $\delta\to1$ which is a critical value that the dispersive bands on either side of the flat band change the direction of their bending, the correlation length has asymptotic values and they are different for s-and d-wave pairing especially at strong-coupling limit. It should be noted that the divergence as 
$\delta\to 0$ is unphysical because the isolated-flat-band condition required by our projection theory is not met in this regime. In particular, at $\delta=0$, gap-closing points occur at the corners of the Brillouin zone. The quantum geometric third-harmonic current is shown in Fig. \ref{fig:Results}(d). It is mainly contributed by the Higgs-mode-mediated fluctuations with the contribution from pure quasiparticle fluctuations being approximately $10^5$ times smaller and thus nearly negligible. The divergence at $\Omega=\Delta_0$ manifests the resonance of Higgs-mode.

\textit{Discussion and conclusion}.---In summary, we analytically calculate two  observables for the Higgs mode in flat-band superconductors: the quantum geometric contribution for the correlation length and the optical kernel related directly to the current. We demonstrate that the third-harmonic electromagnetic response remains present in the flat band and is entirely attributable to the local band geometry. Specifically, the quantum metric comes into play at the interaction vertices between the gauge field and the quasi-particle bubble. Furthermore, we calculate these two quantities for the Lieb lattice model finding that the correlation lengths associated with the extended s-wave and d-wave pairing symmetries show no difference across various coupling strengths. In contrast, the simplest form of s-wave pairing differs markedly from the former two cases, particularly in the strong-coupling regime. Additionally, we find that Higgs-mode-mediated fluctuations dominate the optical kernel and the corresponding third-harmonic current. This differs from single-band systems\cite{THG1}, where the contribution of Higgs-mode fluctuations as an intermediate process is generally overshadowed by quasiparticles. Our results thus reveal a strong dependence of the third-harmonic response on Higgs-mode fluctuations in flat-band systems. 

These conclusions, combined with the fact that all the quantum geometric effects we proposed are fundamental and robust against extrinsic perturbations, imply that detecting Higgs modes in flat-band systems can yield purer and more distinct results, providing a clear signature of Higgs mode in superconductors. 

\begin{acknowledgments}
This work was financially
supported by the National Key R\&D Program of China (Grants No.
2022YFA1403200), National Natural Science Foundation of
China (Grants No. 92265104, No. 12022413, No. 11674331), the Basic Research Program of the Chinese
Academy of Sciences Based on Major Scientific Infrastructures (Grant No. JZHKYPT-2021-08), the CASHIPS Director’s Fund (Grant No. BJPY2023A09), the \textquotedblleft
Strategic Priority Research Program (B)\textquotedblright\ of the Chinese
Academy of Sciences (Grant No. XDB33030100), Anhui Provincial Major S\&T Project (Grant No. s202305a12020005), and  the Major Basic Program of Natural
Science Foundation of Shandong Province (Grant No. ZR2021ZD01), and the High Magnetic Field Laboratory of Anhui Province under Contract
No. AHHM-FX-2020-02.
\end{acknowledgments}

\bibliography{ref}

\end{document}